\documentclass[titlepage,twoside,12pt]{article}
\usepackage{amssymb}
\usepackage{amsfonts}
\begin{document}

{\bf \Large No-Cloning in Reduced Power \\ \\ Algebras} \\ \\

{\bf Elem\'{e}r E Rosinger} \\ \\
Department of Mathematics \\
and Applied Mathematics \\
University of Pretoria \\
Pretoria \\
0002 South Africa \\
eerosinger@hotmail.com \\ \\

{\bf Abstract} \\

The No-Cloning property in Quantum Computation is known not to depend on the unitarity of the operators involved, but only
on their linearity. Based on that fact, here it is shown that the No-Cloning property remains valid when Quantum Mechanics
is re-formulated within far wider frameworks of {\it scalars}, namely, one or the other of the infinitely many {\it
reduced power algebras} which can replace the usual real numbers $\mathbb{R}$, or complex numbers $\mathbb{C}$. \\ \\

{\bf 1. Preliminaries} \\

A remarkable feature of the so called No-Cloning property in Quantum Computation, [3,2], is that it is but a rather
elementary and direct consequence of the {\it linearity} property of unitary operators on finite dimensional complex
Hilbert spaces, and in fact, it does {\it not} require that the respective operators by unitary. The fact that unitary
operators are involved in Quantum Computation is natural and unavoidable, since in Quantum Mechanics it is axiomatic that
the evolution of a quantum systems which is not under measurement is given by such operators, being described by the
Schr\"{o}dinger equation. \\

Based on the above elementary fact underlying usual No-Cloning, here we shall show that the No-Cloning property remains
valid when Quantum Mechanics is re-formulated within what appears to be a far more wide and appropriate framework of {\it
scalars}, namely, any one of the infinitely many algebras which belong to the class of {\it reduced power algebras},
[4-12]. \\

One of the essential features of scalars in algebras of reduced powers is that, in addition to being finite, just as the
usual real or complex numbers, such scalars in algebras of reduced powers can also be {\it infinitesimal}, or on the
contrary, {\it infinitely large}. Consequently, vast opportunities for algebraic operations are opened, and also, for
appropriate physical interpretations. For instance, one may consider the possibility that the Planck constant $h$ is a
nonzero positive infinitesimal, and/or the speed of light $c$ is positive and infinitely large, [12, section 4]. \\ \\

{\bf 2. An Extension of the No-Cloning Property} \\

We recall that the field $\mathbb{R}$ of usual real numbers can be extended into any of the infinitely many possible so
called {\it reduced power algebras} $\mathbb{R}_{\cal F}$, where ${\cal F}$ suitable filters on the set $\mathbb{N}$ of
natural numbers, see Appendix, and for further details [12, pp. 3-6], [4-11]. Similarly, the field $\mathbb{C}$ of usual complex
numbers can be extended into any of the infinitely many possible {\it reduced power algebras} $\mathbb{C}_{\cal F}$.
Furthermore, some of these algebras $\mathbb{R}_{\cal F}$ and $\mathbb{C}_{\cal F}$ are themselves fields, namely, when
${\cal F}$ are ultrafilters on the set $\mathbb{N}$ of natural numbers. \\

Let us now recall that in usual Quantum Computation, a quantum register of one qubit is represented as a vector in the
complex Hilbert space $\mathbb{C}^2$. And in general, a quantum register of $n \geq 1$ qubits is represented by the
$n$-fold tensor product \\

(2.1)~~~ $ H_n ~=~ \mathbb{C}^2 \otimes \ldots \otimes \mathbb{C}^2 \thickapprox \mathbb{C}^{2^n} $ \\

Here, we shall replace such usual quantum registers of $n \geq 1$ qubits by the larger spaces \\

(2.2)~~~ $ H_{{\cal F},\, n} ~=~ ( \mathbb{C}_{\cal F} )^2 \otimes \ldots \otimes ( \mathbb{C}_{\cal F} )^2 $ \\

with $n \geq 1$ factors, which again are vector spaces over $\mathbb{C}$, as can easily be seen in [12, pp. 3-6], [4-11]. Furthermore, they possess
an extended scalar product (A.20) - (A.26) which gives them properties similar with the usual Hilbert spaces, properties
sufficient in order to establish the extended version of the No-Cloning property. \\

What is important to note is that, since the No-Cloning property does {\it not} in fact require the unitarity of the
operators involved, but only their {\it linearity}, we can proceed with the extension of the No-Cloning property to
quantum registers given by the vector spaces $H_{{\cal F},\, n}$ in (2.2), without having to consider on them any usual
Hilbert space structure, and instead, by only using the above mentioned extended Hilbert space structure of these
spaces. \\

In order to make clear this argument, let us briefly recall the usual No-Cloning property, [3]. \\

First, let us note that, scientists are on occasion giving names to new phenomena in ways which are not thoroughly well
considered, and thus may lend themselves to misinterpretation. One such case is, unfortunately, with the term {\it
No-Cloning} used in Quantum Computation. What is in fact going on here is that, quite surprisingly, quantum computers do
{\it not} allow the copying of {\it arbitrary} qubits. And here by "copying" one means the precise reproduction any finite
number of times of a given arbitrary qubit, a reproduction which does {\it not} destroy the original qubit which is being
reproduced. \\
Thus a more proper term would be the somewhat longer one of {\it no arbitrary copying}. \\
Yet in spite of that, plenty of copying can be done by quantum computers, as will be seen in the sequel. \\

In order better to understand the issue, let us start by considering copying classical bits. For that purpose we can use
the classical version of the quantum CNOT gate, [2,3], operating this time on bits $a,~ b \in \{~ 0, 1 ~\}$, namely

\begin{math}
\setlength{\unitlength}{1cm}
\thicklines
\begin{picture}(15,4)

\put(2.8,3.8){$\mbox{a}$}
\put(3.5,3.9){\line(1,0){5}}
\put(9,3.8){$\mbox{a}$}
\put(2.8,1.1){$\mbox{b}$}
\put(3.5,1.2){\line(1,0){5}}
\put(9,1.1){$\mbox{a} ~\oplus~ \mbox{b}$}
\put(6,3.85){\circle*{0.5}}
\put(6,1.2){\line(0,1){2.6}}
\put(6,1.2){\circle{0.5}}
\put(5.1,0){$\mbox{Fig. 2.1.}$}

\end{picture}
\end{math}

\bigskip

Now, if we fix $b = 0$, then for an arbitrary input bit $a \in  \{~ 0, 1 ~\}$, we shall obtain as output two copies of
$a$. \\

Strangely enough, a similar copying of arbitrary quantum bits cannot be performed by quantum systems, as was discovered in
1982 by W K Wooters and W H Zurek, [2,3]. \\
Of course, as well known, [2,3], each qubit contains a double infinity of classical information since it can be an
arbitrary point on the Bloch sphere, which is much unlike the situation with one single bit. In this way, the ability to
copy arbitrary qubits is considerably more demanding than copying arbitrary classical bits. \\

Let us now turn to this issue in some more detail. First we present a simple and somewhat intuitive argument. We assume
that we have a quantum system $S$ which allows one qubit at input and has one qubit at output. The output facility we
shall use as a "blank sheet" on which we want to copy an arbitrary input qubit $|~ \psi > ~\in \mathbb{C}^2$. We can
assume that the initial state of the "blank sheet" at the output is given by a fixed qubit $|~ \chi_0 > ~\in \mathbb{C}^2$.
Thus we start with the setup

\begin{math}
\setlength{\unitlength}{1cm}
\thicklines
\begin{picture}(15,4.5)

\put(2,1.9){$|~ \psi >$}
\put(3.5,2){\line(1,0){1.5}}
\put(5,1){\line(0,1){2}}
\put(5,1){\line(1,0){2}}
\put(5,3){\line(1,0){2}}
\put(7,1){\line(0,1){2}}
\put(7,2){\line(1,0){1.5}}
\put(9,1.9){$|~ \chi_0 >$}
\put(5.8,1.9){$S$}
\put(5.1,0){$\mbox{Fig. 2.2.}$}

\end{picture}
\end{math}

\bigskip

and would like to end up with the setup

\begin{math}
\setlength{\unitlength}{1cm}
\thicklines
\begin{picture}(15,4)

\put(2,1.9){$|~ \psi >$}
\put(3.5,2){\line(1,0){1.5}}
\put(5,1){\line(0,1){2}}
\put(5,1){\line(1,0){2}}
\put(5,3){\line(1,0){2}}
\put(7,1){\line(0,1){2}}
\put(7,2){\line(1,0){1.5}}
\put(9,1.9){$|~ \psi >$}
\put(5.8,1.9){$S$}
\put(5.1,0){$\mbox{Fig. 2.3.}$}

\end{picture}
\end{math}

\bigskip

However, as quantum processes evolve through unitary operators when not subjected to measurement, it means that we are
looking for such a unitary operator $U : \mathbb{C}^2 ~\otimes~ \mathbb{C}^2 ~\longrightarrow~ \mathbb{C}^2 ~\otimes~
\mathbb{C}^2$, and one which would act according to \\

(2.3)~~~ $ U (~ |~ \psi > ~\otimes~ |~ \chi_0 > ~) ~=~ |~ \psi > ~\otimes~ |~ \psi >,~~~~ |~ \psi > ~\in \mathbb{C}^2 $ \\

Before going further, let us immediately remark here that a unitary operator $U$, which therefore is linear, is not likely
to satisfy (2.3), in view of the fact that this is a {\it nonlinear} relation in  $|~ \psi > ~\in \mathbb{C}^2$, and in particular, its left hand term
is linear in  $|~ \psi >$, while its right hand term is a quadratic in  $|~ \psi >$. \\

And now, let us return to a more precise argument. Since $|~ \psi > ~\in \mathbb{C}^2$ is assumed to be arbitrary in (2.3),
we can write that relation for any $|~ \psi_1 >,~|~ \psi_2 > ~\in \mathbb{C}^2$. Thus we obtain \\

(2.4)~~~ $ \begin{array}{l}
                U (~ |~ \psi_1 > ~\otimes~ |~ \chi_0 > ~) ~=~ |~ \psi_1 > ~\otimes~ |~ \psi_1 > \\ \\
                U (~ |~ \psi_2 > ~\otimes~ |~ \chi_0 > ~) ~=~ |~ \psi_2 > ~\otimes~ |~ \psi_2 >
           \end{array} $ \\

Now if we take the inner product of these two relations and recall that $U$ was supposed to be unitary, we obtain \\

(2.5)~~~ $ < \psi_1~ ~|~  \psi_2 > ~=~ (~ < \psi_1 ~|~  \psi_2 > ~)^2 $ \\

which implies that either $< \psi_1~ ~|~  \psi_2 > ~=~ 0$, or $< \psi_1~ ~|~  \psi_2 > ~=~ 1$. This means that the two
arbitrary quantum states $|~ \psi_1 >,~ |~ \psi_2 > ~\in \mathbb{C}^2$ are always either orthogonal, or identical from
quantum point of view, which is clearly absurd. \\

The general and rigorous argument is as follows. We consider a quantum system whose state space is $\mathbb{C}^m$, for a
certain integer $m \geq 2$. Further, we fix in this state space an arbitrary orthonormal basis $|~ \psi_1 >,~ .~.~.~ ,
|~ \psi_m > ~\in \mathbb{C}^m$. Finally, we assume that the state $|~ \psi_1 >~$ will function as the "blank sheet" on
which we want to copy arbitrary states $|~ \psi > ~\in \mathbb{C}^m$. \\

Then the desired copying machine of arbitrary states in $\mathbb{C}^m$ will be given by a unitary operator $U :
\mathbb{C}^m ~\otimes~ \mathbb{C}^m ~\longrightarrow~ \mathbb{C}^m ~\otimes~ \mathbb{C}^m$, for which we have \\

(2.6)~~~ $ U (~ |~ \psi > ~\otimes~ |~ \psi_1 > ~) ~=~ |~ \psi > ~\otimes~ |~ \psi >,~~~~ |~ \psi > ~\in \mathbb{C}^m $ \\

And now we can prove that for $n \geq 2$, there does {\it not} exist such a copying machine $U$. \\

Indeed, if we assume that $n \geq 2$, then we do have at least the two orthonormal states $|~ \psi_1 >,~ |~ \psi_2 > ~\in
\mathbb{C}^m$. Thus (2.6) gives \\

(2.7)~~~ $ \begin{array}{l}
                   U (~ |~ \psi_1 > ~\otimes~ |~ \psi_1 > ~) ~=~ |~ \psi_1 > ~\otimes~ |~ \psi_1 > \\ \\
                   U (~ |~ \psi_2 > ~\otimes~ |~ \psi_1 > ~) ~=~ |~ \psi_2 > ~\otimes~ |~ \psi_2 > \\ \\
                   U  (~ (~ |~ \psi_1 > ~+~ |~ \psi_2 > ~) ~\otimes~ |~ \psi_1 > ~) ~=~ \\ \\
                      ~~~~~=~ (~ |~ \psi_1 > ~+~ |~ \psi_2 > ~) ~\otimes~ (~ |~ \psi_1 > ~+~ |~ \psi_2 > ~)
           \end{array} $ \\ \\

Now the linearity of $U$ gives together with the first two relations above \\

(2.8)~~~ $ \begin{array}{l}
                   U  (~ (~ |~ \psi_1 > ~+~ |~ \psi_2 > ~) ~\otimes~ |~ \psi_1 > ~) ~=~ \\ \\
                     ~~~~~ =~ U (~ |~ \psi_1 > ~\otimes~ |~ \psi_1 > ~) ~+~ U (~ |~ \psi_2 > ~\otimes~
                     |~ \psi_1 > ~) ~=~ \\ \\
                     ~~~~~ =~ |~ \psi_1 > ~\otimes |~ \psi_1 > ~+~ |~ \psi_2 > ~\otimes |~ \psi_2 >
                         \end{array} $ \\

Thus (2.8) with the last relation in (2.7) imply \\

(2.9)~~~ $ \begin{array}{l}
                  (~ |~ \psi_1 > ~+~ |~ \psi_2 > ~) ~\otimes~ (~ |~ \psi_1 > ~+~ |~ \psi_2 > ~) ~=~ \\ \\
                             ~~~~~~~~~~~~~~~~~~~~~~~ =~ |~ \psi_1 > ~\otimes |~ \psi_1 > ~+~ |~ \psi_2 > ~\otimes |~ \psi_2 >
           \end{array} $ \\

or in other words \\

(2.10)~~~ $ |~ \psi_1 > ~\otimes~ |~ \psi_2 > ~+~ |~ \psi_2 > ~\otimes~ |~ \psi_1 > ~=~ 0 $ \\

which is obviously false. \\

Let us point out two facts with respect to the above no-cloning result. \\

First, in the more general second proof, we did {\it not} use the fact that $U$ is unitary, and only made use of its
linearity, when we obtained (2.8). In the first proof, on the other hand, the fact that $U$ is unitary was essential in
order to obtain (2.5). \\

Second, it is important to understand properly the meaning of the above limitation implied by No-Cloning. Indeed, while it
clearly does not allow the copying of arbitrary qubits, it does nevertheless allow the copying of a {\it large range} of
qubits. \\
For instance, in terms of the second proof, let $I = \{~ 1, ~.~.~.~ , n ~\}$ be the set of indices of the respective
orthonormal basis \\

$~~~~~~ |~ \psi_1 >,~ .~.~.~ , |~ \psi_n > ~\in \mathbb{C}^m $ \\

Further, let us consider the partially defined function \\

$~~~~~~ c : I \times I ~\longrightarrow~ I \times I $ \\

given by $c~ ( i, 1 ) = ( i, i )$, with $1 \leq i \leq n$. Then clearly, $c$ is injective on the domain on which it is
defined. Therefore, $c$ can be extended to the whole of $I \times I$, so as still to remain injective, and in fact, to
become bijective as well. And obviously, there are many such extensions when $n \geq 2$. \\

Now we can define a mapping $U$ by \\

$~~~~~~ U (~ |~ \psi_i > ~\otimes~ |~ \psi_j > ~) ~=~ |~ \psi_k > ~\otimes~ |~ \psi_l > $ \\

where $1 \leq i,~ j \leq n$ and $c~ ( i, j ) = ( k, l )$. Since $c$ is bijective on $I \times I$, this mapping $U$ will be
a permutation of the respective basis in $\mathbb{C}^m ~\otimes~ \mathbb{C}^m$, therefore it extends in a unique manner to
a linear and unitary mapping \\

$~~~~~~ U : \mathbb{C}^m ~\otimes~ \mathbb{C}^m ~\longrightarrow~ \mathbb{C}^m ~\otimes~ \mathbb{C}^m $ \\

And now it follows that \\

$~~~~~~ U (~ |~ \psi_i > ~\otimes~ |~ \psi_1 > ~) ~=~ |~ \psi_i> ~\otimes~ |~ \psi_i >,~~~~~ 1 \leq i \leq n $ \\

thus indeed $U$ is a copying machine with the "blank sheet" $|~ \psi_1 >$, and it can copy onto this "blank sheet" {\it
all} the qubits in the given orthonormal basis $|~ \psi_1 >,~ .~.~.~ , |~ \psi_n >~$ of $\mathbb{C}^m$. And in any such
basis, with the exception of the fixed "blank sheet" $|~ \psi_1 >$, all the other qubits $|~ \psi_2 >,~ .~.~.~ ,
|~ \psi_n >~$ are {\it arbitrary}, within the constraint that together they have to form an orthonormal basis. \\

Returning now to the extended situation in (2.2), we obtain the following No-Cloning property \\

{\bf Theorem 2.1. ( Extended No-Cloning )} \\

Given any extended quantum register  $H_{{\cal F},\, n}$, and $\psi_1, \ldots , \psi_m ~\in ( \mathbb{C}_{\cal F} )^m$ orthonormal vectors, where $n, m
\geq 2$. Then there does {\it not} exist any linear operator \\

(2.11)~~~ $ U : ( \mathbb{C}_{\cal F} )^m ~\otimes~ ( \mathbb{C_{\cal F}} )^m ~\longrightarrow~
                        ( \mathbb{C}_{\cal F} )^m ~\otimes~ ( \mathbb{C}_{\cal F} )^m $ \\

such that \\

(2.12)~~~ $ U ( \psi ~\otimes~ \psi_1 ~) ~=~ \psi ~\otimes~ \psi,~~~~ \psi \in \mathbb{C}^m $ \\

{\bf Proof.} \\

We note that the relations (2.7) - (2.10) extend easily to (2.11), (2.12). \\ \\

{\bf Appendix} \\

For convenience, we shall recall in a particular case the construction, [4-11], as reviewed in [12, pp. 3-6], of {\it reduced power algebras}. Given any
{\it filter} ${\cal F}$ on $\mathbb{N}$, we define \\

(A.1)~~~ $ {\cal I}_{\cal F} = \{~ v = ( v_n )_{n \in \mathbb{N}} \in \mathbb{R}^\mathbb{N} ~~|~~
                                       \{~ n \in \mathbb{N} ~|~ v_n = 0 ~\} \in {\cal F} ~\} $ \\

which is a {\it proper ideal} in the algebra $\mathbb{R}^\mathbb{N}$. Thus we obtain the {\it reduced power algebra} associated to ${\cal F}$ as the quotient algebra \\

(A.2)~~~ $ \mathbb{R}_{\cal F} = \mathbb{R}^\mathbb{N} / {\cal I}_{\cal F} $ \\

Furthermore, this algebra which is commutative, is also a strict extension of the field $\mathbb{R}$ of the usual real numbers, according to the embedding of algebras \\

(A.3)~~~ $ \mathbb{R} \ni x \longmapsto ( x, x, x, \ldots ) + {\cal I}_{\cal F} \in \mathbb{R}_{\cal F} = \mathbb{R}^\mathbb{N} / {\cal I}_{\cal F} $ \\

In a similar manner one can obtain reduced power algebras extending the field $\mathbb{C}$ of the usual complex numbers. Namely, let us denote by \\

(A.4)~~~ $ {\cal J}_{\cal F} = \{~ w = ( w_n )_{n \in \mathbb{N}} \in \mathbb{C}^\mathbb{N} ~~|~~
                                       \{~ n \in \mathbb{N} ~|~ w_n = 0 ~\} \in {\cal F} ~\} $ \\

which is a {\it proper ideal} in the algebra $\mathbb{C}^\mathbb{N}$. Thus we obtain the {\it reduced power algebra} associated to ${\cal F}$ as the quotient algebra \\

(A.5)~~~ $ \mathbb{C}_{\cal F} = \mathbb{C}^\mathbb{N} / {\cal J}_{\cal F} $ \\

Furthermore, this algebra which is commutative, is also a strict extension of the field $\mathbb{C}$ of the usual complex numbers, according to the embedding of algebras \\

(A.6)~~~ $ \mathbb{C} \ni z \longmapsto ( z, z, z, \ldots ) + {\cal J}_{\cal F} \in \mathbb{C}_{\cal F} =
                             \mathbb{C}^\mathbb{N} / {\cal J}_{\cal F} $ \\

We now establish a natural connection between the algebras $\mathbb{R}_{\cal F}$ and $\mathbb{C}_{\cal F}$. \\

In this regard, we note the following connection between the ideals ${\cal I}_{\cal F}$ and ${\cal J}_{\cal F}$. Namely \\

(A.7)~~~ $ \begin{array}{l}
                  w = ( w_n = u_n + i v_n )_{n \in \mathbb{N}} \in {\cal J}_{\cal F} ~~\Longleftrightarrow \\ \\
                  ~~~~~~~\Longleftrightarrow~~ u = ( u_n )_{n \in \mathbb{N}},~
                             v = ( v_n )_{n \in \mathbb{N}} \in {\cal I}_{\cal F}
           \end{array} $ \\

where $u_n, v_n \in \mathbb{R}$. It follows that we have the algebra homomorphisms \\

(A.8)~~~ $ \begin{array}{l}
              Re : \mathbb{C}_{\cal F} \ni  w = ( w_n = u_n + i v_n )_{n \in \mathbb{N}} + {\cal J}_{\cal F} \longmapsto \\ \\
              ~~~~~~ \longmapsto
                 u = ( u_n )_{n \in \mathbb{N}} + {\cal I}_{\cal F} \in \mathbb{R}_{\cal F}
           \end{array} $ \\ \\

(A.9)~~~ $ \begin{array}{l}
               Im : \mathbb{C}_{\cal F} \ni  w = ( w_n = u_n + i v_n )_{n \in \mathbb{N}} + {\cal J}_{\cal F} \longmapsto \\ \\
               ~~~~~~ \longmapsto
                 v = ( v_n )_{n \in \mathbb{N}} + {\cal I}_{\cal F} \in \mathbb{R}_{\cal F}
           \end{array} $ \\

as well as the algebra embeddings \\

(A.10)~~~ $ \mathbb{R}_{\cal F} \ni u = ( u_n )_{n \in \mathbb{N}} + {\cal I}_{\cal F} \longmapsto
                            u = ( u_n )_{n \in \mathbb{N}} + {\cal J}_{\cal F} \in \mathbb{C}_{\cal F} $ \\

(A.11)~~~ $ \mathbb{R}_{\cal F} \ni v = ( v_n )_{n \in \mathbb{N}} + {\cal I}_{\cal F} \longmapsto
                            i v = ( i v_n )_{n \in \mathbb{N}} + {\cal J}_{\cal F} \in \mathbb{C}_{\cal F} $ \\

Let us also define the surjective linear mapping \\

(A.12)~~~ $ \begin{array}{l}
                 \mathbb{C}_{\cal F} \ni  w = ( w_n = u_n + i v_n )_{n \in \mathbb{N}} + {\cal J}_{\cal F} \longmapsto \\ \\
                 ~~~~~~ \longmapsto
                 \overline{w} = ( \overline{w_n} = u_n - i v_n )_{n \in \mathbb{N}} + {\cal J}_{\cal F} \in \mathbb{C}_{\cal F}
             \end{array} $ \\

As a consequence, we obtain \\

(A.13)~~~ $ w = ( w_n = u_n + i v_n )_{n \in \mathbb{N}} + {\cal J}_{\cal F} \in \mathbb{C}_{\cal F},~~~
                  \overline{w} = w
           ~~~\Longrightarrow~~~ w \in \mathbb{R}_{\cal F} $ \\

Lastly, we can define the {\it absolute value} on $\mathbb{C}_{\cal F}$, by the mapping \\

(A.14)~~~ $ \begin{array}{l}
                \mathbb{C}_{\cal F} \ni z = ( w_n = u_n + i v_n )_{n \in \mathbb{N}} + {\cal J}_{\cal F}
                                 \longmapsto \\ \\
                ~~~~~~ \longmapsto
                   | z | = ( | w_n | = \sqrt ( u^2_n + v^2_n ) )_{n \in \mathbb{N}} + {\cal I}_{\cal F} \in \mathbb{R}_{\cal F}
            \end{array} $ \\

Let us denote \\

(A.15)~~~ $ \mathbb{R}^+_{\cal F} = \{~ u = ( u_n )_{n \in \mathbb{N}} + {\cal I}_{\cal F} \in \mathbb{R}_{\cal F}
                     ~~|~~ \{~ n \in \mathbb{N} ~|~ u_n \geq 0 ~\} \in {\cal F} ~\} $ \\

then we obtain the surjective mapping \\

(A.16)~~~ $  \mathbb{C}_{\cal F} \ni z \longmapsto | z | \in \mathbb{R}^+_{\cal F} $ \\

and for $z \in \mathbb{C}_{\cal F}$, we have \\

(A.17)~~~ $ | z | = 0 ~~\Longleftrightarrow~~ z = 0 $ \\

Now, in view of (A.8), (A.9), (A.14), we have for $z \in \mathbb{C}_{\cal F}$ the relations \\

(A.18)~~~ $ |\, Re \, z \,|,~~ |\, Im \, z \,| ~\leq~ |\, z \,| $ \\

where the partial order $\leq$ is defined on $\mathbb{C}_{\cal F}$ by \\

(A.19)~~~ $ u \leq v ~~~ \Longleftrightarrow~~~ v - u \in \mathbb{R}^+_{\cal F} \\$

Lastly, for $m \geq 1$, we define an {\it extended scalar product} \\

(A.20)~~~ $ < , > \, : ( \mathbb{C}_{\cal F} )^m \times ( \mathbb{C}_{\cal F} )^m ~\longrightarrow~ \mathbb{C}_{\cal F} $ \\

by \\

(A.21)~~~ $ < ( z_1, \ldots , z_m ), ( w_1, \ldots , w_m ) > ~=~ \overline{z_1} w_1 + \ldots + \overline{z_m} w_m \in \mathbb{C}_{\cal F} $ \\

for $\psi = ( z_1, \ldots , z_m ),~ \chi = ( w_1, \ldots , w_m ) \in ( \mathbb{C}_{\cal F} )^m$. \\

Then this extended scalar product has the properties \\

(A.22)~~~ It is linear over $\mathbb{C}_{\cal F}$, therefore also over $\mathbb{C}$, in the second \\
         \hspace*{1.3cm} argument. \\

(A.23)~~~ $  < \chi, \psi > ~=~ \overline{< \psi, \chi >},~~~ \psi, \chi \in ( \mathbb{C}_{\cal F} )^m $ \\

(A.24)~~~ $ < \psi, \psi > \,\, \in \mathbb{R}^+_{\cal F},~~~ \psi \in ( \mathbb{C}_{\cal F} )^m $ \\

and for $\psi \in ( \mathbb{C}_{\cal F} )^m$, one has \\

(A.25)~~~ $ < \psi, \psi > ~=~ 0 ~~\Longleftrightarrow~~ \psi = 0 \in ( \mathbb{C}_{\cal F} )^m $ \\

Also, we have the extension of the classical Schwartz inequality \\

(A.26)~~~ $ |\, < \psi, \chi > \,\,| ~\leq~ < \psi, \psi >^{1/2} \,\, < \chi, \chi >^{1/2},~~~
                                  \psi, \chi \in ( \mathbb{C}_{\cal F} )^m $ \\

Two vectors $\psi, \chi \in ( \mathbb{C}_{\cal F} )^m$ are called {\it orthogonal}, if and only if $< \psi, \chi > ~=~ 0$. \\

Two orthogonal vectors $\psi, \chi \in ( \mathbb{C}_{\cal F} )^m$ are called {\it orthonormnal}, if and only if $< \psi, \psi > ~=~ < \chi, \chi > ~=~
1$. \\


\begin{thebibliography}{99}

\bibitem{} Gillespie D T : A Quantum Mechanics Primer, An Elementary Inroduction to the Formal Theory of Nonrelativistic
Quantum Mechanics. Open University Set Book, International Textbook Company Ltd., 1973, ISBN 0 7002 2290 1

\bibitem{} Nielsen M A, Chuang I L : Quantum Computation and Quantum Information. Cambridge Univ. Press, 2000

\bibitem{} Rosinger E E : Basics of Quantum Computation (Part I) \\ arxiv:quant-ph/0407064

\bibitem{} Rosinger E E : What scalars should we use ? \\ arXiv:math/0505336

\bibitem{} Rosinger E E : Solving Problems in Scalar Algebras of Reduced Powers. arXiv:math/0508471

\bibitem{} Rosinger E E : From Reference Frame Relativity to Relativity of Mathematical Models :
Relativity Formulas in a Variety of non-Archimedean Setups. arXiv:physics/0701117

\bibitem{} Rosinger E E : Cosmic Contact : To Be, or Not To Be \\ Archimedean ? arXiv:physics/0702206

\bibitem{} Rosinger E E : String Theory: a mere prelude to \\ non-Archimedean Space-Time Structures? \\
arXiv:physics/0703154

\bibitem{} Rosinger E E : Mathematics and "The Trouble with Physics", How Deep We Have to Go ? arXiv:0707.1163

\bibitem{} Rosinger E E : How Far Should the Principle of Relativity Go ? arXiv:0710.0226

\bibitem{} Rosinger E E : Archimedean Type Conditions in Categories. arXiv:0803.0812

\bibitem{} Rosinger E E : Heisenberg Uncertainty in Reduced Power Algebras. arxiv:0901.4825

\end{thebibliography}
\end{document}